\documentclass[<options>]{elsarticle}
\usepackage[utf8]{inputenc}
\usepackage{mathrsfs}
\usepackage{siunitx}
\usepackage{subcaption} 

\title{A Hilbert Transform method for measuring linear and nonlinear phase shifts imparted by metasurfaces }

\author[1]{Laura C. Wynne}

\author[1]{Hamish T. Ballantyne}

\author[1]{Xin Li}

\author[1]{Andrea di Falco}

\author[1]{Sebastian A. Schulz}

\address[1]{SUPA, School of Physics and Astronomy, University of St Andrews, North Haugh, St Andrews KY16 9SS, United Kingdom }

\begin{document}

\begin{abstract} 

Nonlinear metasurfaces that dynamically manipulate the phase of a passing light beam are of interest for a wide range of applications. The controlled operation of such devices requires accurate measurements of the optical transmission phase in both the linear and nonlinear regime, an experimentally challenging task. In this paper we show that this phase information can be extracted directly from simple transmission measurements, using a Hilbert transform approach, removing the need for complicated, interferometric experimental set-ups, and enabling direct measurements of the phase in conditions not suitable for other traditional approaches, such Z-scan measurements.

\end{abstract}

\maketitle

\section{Introduction}
Many metasurfaces function by constructing a spatial phase gradient which allows the shaping of incident wavefronts, creating devices such as meta-lenses, phaseplates or holograms \cite{Khorasaninejadeaam8100,Yu2011,doi:10.4218/etrij.2018-0532}. Recent advances have led to the realization of new classes of metasurfaces that allow dynamic control of the incident phase, through mechanical, thermal, or nonlinear optical manipulations \cite{doi:10.1021/acs.nanolett.6b00618,doi:10.1021/acsphotonics.7b01517,Alam2018}. All of these applications require knowledge of the imparted phase and potentially dynamic variations in the phase, to design meta-atoms or antenna arrangements. Therefore there is a need for accurate phase retrieval methods, especially relevant to practical  experimental characterization techniques. A further condition is that these phase retrieval methods should be capable of addressing both the linear and nonlinear phase of the metasurfaces, both experimentally and during the design stage. Methods based on simple and quick computation are especially desirable for the robust design of metasurfaces. 

Traditionally, the phase of such metasurfaces can be extracted from both simulation (i.e. the design phase) and experiments, i.e. the realised phase. Experimental methods of determining the phase of an optical device are typically based on interferometric measurements, such as Fourier Transform Spectral Interferometry (FTSI). The working of this method is discussed in more detail below, however we here quickly note that optical metasurfaces are typically designed to provide a phase delay between 0 and $2\pi$. These very small delays are easily overshadowed by other, larger variations in an interferometric set-up, such as vibrations or temperature fluctuations. Furthermore, the FTSI method cannot easily be applied to the antenna design stage, which is typically performed using numerical methods, such as Finite Difference timed-Domain (FDTD) or finite element methods (FEM). Both approaches are computationally expensive, precluding the implementation of interferometers in the simulation step. Instead, at the design step, the phase can be calculated in a brute force manner. For example, in FDTD the 2-dimensional field profile of the meta-atom/metasurface can be recorded in the far-field and the phase extracted from its complex component. However, this  far-field propagation is very time consuming, and, as we will show, unnecessary. 
Instead, we show that Hilbert transformations \cite{Sima2013} can be used to extract both the linear and nonlinear phase, eliminating the need for interferometric set-ups or Z-scan measurements. The linear phase is extracted from a single, normalised transmission spectrum and the nonlinear phase shift from two steady state transmission spectra. This Hilbert transform method is applicable to both numerical and experimental transmission data, removing the need for computational propagation into the far-field, or interferometric set-ups.

\begin{figure}
    \centering
    \includegraphics[scale=0.4]{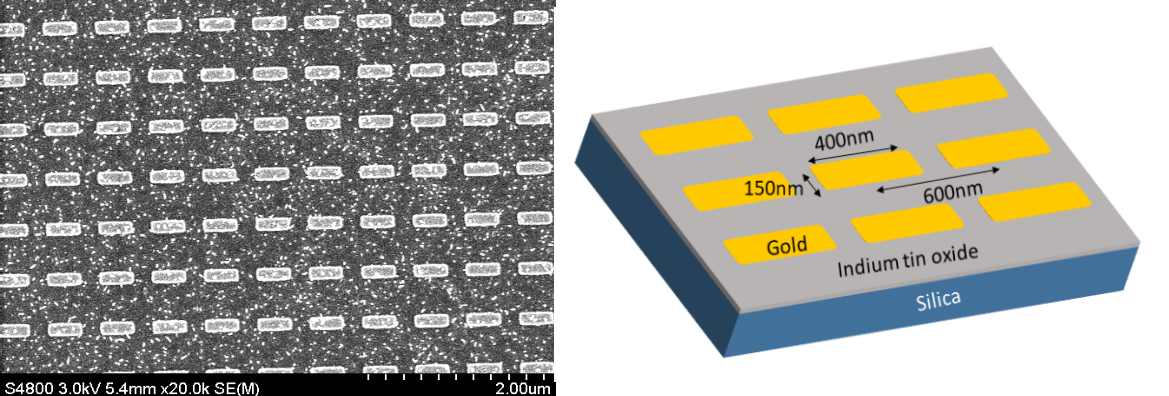}
    \caption{Left panel: Scanning electron microscope (SEM) image of the gold antenna array fabricated on top of an ITO thin film, Right panel: Schematic of the metasurface, showing both the periodic antenna array and the layered device structure.}
    \label{fig:2}
\end{figure}

As an exemplary metasurface we are considering plasmonic antennas on an epsilon-near-zero (ENZ) film. These structures have recently been shown to provide coupling between the modes of the thin ENZ film and the nanoantennas \cite{PhysRevA.93.063846}, and also to show a strong nonlinear response suitable for dynamic modulation of the metasurfaces \cite{Guo2016,Alam2018}. For this work, the metasurface consists of a thin film of indium tin oxide (ITO) sputtered onto a glass slide, upon which a periodic array of gold antennas has been fabricated via e-beam lithography,  using a PMMA resist bi-layer (exposed at 30kV acceleration voltage), gold evaporation and lift-off. The ITO was sputtered using an RF power source in a 20 SCCM Argon environment at a pressure of 3 mTorr, at a power of 60 W, at \SI{200}{\celsius} with a half hour post deposition annealing time, resulting in an epsilon-zero wavelength in the infrared range (approximately 1470 nm). The gold is deposited using electron beam evaporation. Both the gold and ITO thin films are 40 nm in thickness, with the nominal antenna dimensions being width 150 nm, length 400 nm and period 600 nm. Additionally, intra-structure proximity error correction has been applied to achieve smaller corner radii \cite{Schulz2015}. For these thicknesses the ITO thin-film supports ENZ modes \cite{Campione2015} that couple to the antenna modes \cite{PhysRevA.93.063846}. A sketch of the sample geometry and a typical SEM picture of a fabricated sample are shown in fig.\ref{fig:2}.

  %

\section{Limitations of FTSI}

FTSI has been a common choice for measuring the phase of optical elements, including the phase which a metamaterial can impart to incident light \cite{OBrien:12}, with the method covered in detail in a wide range of papers, e.g. references \cite{Lepetit:95,Pshenay-Severin:10}. We therefore only sketch the method, highlighting the particular challenges of applying it to metasurface devices. A typical FTSI set-up involves recording the interference between two recombined light paths, for example within a Mach-Zehnder or Michelson-Morely interferometer. The device under investigation is placed in one of these arms, from now on referred to as the sample arm, with the other path being the reference arm. The collected interferogram includes an interference term, that holds the information on the delay between the sample and reference arms.
 This term is then Fourier transformed, filtered (e.g. through windowing) and inverse Fourier transformed to isolate the delay. By separately measuring the base delay between the two arms, (in the absence of a sample/device) the sample contribution to the delay can be extracted \cite{doi:10.1063/1.2752761}. As such this technique is relatively straight forward to implement in experiment and can be performed using spectral-domain measurements (rather than time-domain measurements). However, this method typically requires the two arms of the interfereometer to be either perfectly balanced in the absence of a sample, or the imbalance of the two arms to be known. While this can easily be measured through the FTSI techniques, it is then required that the arm imbalance would remain constant throughout subsequent measurements. Any variation, e.g. due to temperature fluctuations, could easily exceed the delay from the sample, in particularly for the very small delays relevant to metasurfaces.
    
In the design of phase gradient metasurfaces one typically has a phase profile that varies between 0 and $2\pi$ Radians, which is challenging to measure experimentally using FTSI. If for example we consider a $\pi$ Radians phase shift which is experienced at all frequencies between \SI{1.1}{\micro \meter} and \SI{1.7}{\micro \meter}, then the corresponding time delay would range from \SI{1.8}{\femto \second} to \SI{2.8}{\femto \second} across the spectrum. Measuring such a small time delay (or the $\lambda/2$) path difference is challenging, as minute changes in the experimental conditions, e.g. vibrations or temperature fluctuations between reference and sample measurement can easily produce variation in the arm imbalance on a much larger scale. E.g. we constructed a typical Mach-Zehnder set-up and balanced the arm length such that the time delay between the sample arm and the reference arm was \SI{0.8}{\pico \second}, corresponding to a physical length difference of only \SI{240}{\micro \meter}. Yet, we were unable to obtain a consistent and useful measurement of the metasurface phase using the FTSI technique (data not shown).

There are several factors which can hinder an accurate measurement. Fabry-P\'erot resonances, originating from unwanted reflections at e.g. fibre facets, lens interfaces or the sample itself can add undesired spectral features in the measurements by introducing additional interferencce fringes and increasing the uncertainty over the phase which is retrieved. Thermal fluctuations and vibrations can also affect the two arms differently and can eclipse the small phase delay that is being measured. Using the Mach-Zehnder interferometer described above, fluctuations which imparted a phase variation of up to 30~rad were observed - far exceeding the relevant metasurface signal. There do exist set-ups which allow the collection of the experimental noise of the sample path, the reference path and the interferogram simultaneously, allowing for more effective noise reduction \cite{Azari:14}. However these set-ups are more involved than that of the original Mach-Zehnder interferometer, adding further experimental complexity set-up. Our aim on the other hand is is to obtain the metasurface's phase response while reducing the complexity of both the experiment and data analysis, through the use of Hilbert transformations.

\section{The Hilbert Transform Method}

Hilbert transforms have important roles in optics, as they relate imaginary and real parts of a function, as exemplified by the well known Kramers-Kronig relations. The Hilbert transform has been a common tool in work concerning communication between classical radio antennas, making it a natural candidate for optical systems. For example, Hilbert transforms have significantly outperformed FTSI in the measurement of group delay in Fibre Bragg gratings \cite{Dicaire:14}. Hilbert transforms have been used frequently for phase retrieval in interferometry as a substitute to FTSI where the Hilbert transform is applied to the interferogram rather than transmission spectrum of a sample \cite{doi:10.1111/j.1475-1305.2008.00451.x, 10.1117/1.OE.55.11.113105}.
While Hilbert transforms for phase retrieval have been applied to metasurfaces before, the method has not been applied to the meta-atom design step, nor the nonlinear optical regime \cite{doi:10.1063/1.4881332}.

Hilbert transforms can be applied to systems which are both linear and causal. If for example a linear function $G(\omega)$ is considered, causality demands that the impulse response must be $g(t)=0$ if $t<0$, where the impulse response is the Fourier transform of $G(\omega)$ \cite{MECOZZI20094183}.The integral however is a divergent integral due to the existence of a singularity when $\omega =\omega_0$. At the point of the singularity the integral is considered to be a Cauchy principal value.


\begin{equation}
    \mathscr{H}[\mathscr{I}[G(\omega)]]=\frac{1}{\pi} \int_{-\infty}^{\infty}\frac{\mathscr{I}[G(\omega)]}{\omega-\omega_0} d\omega =\mathscr{R}[G(\omega)]
\end{equation}

The above equation will shift positive frequency components by -90 degrees and negative ones by 90 degrees. Computing the Hilbert transform of the function $G(\omega)$ is straightforward; the above integral is equivalent to the convolution of $\mathscr{I}[G(\omega)]$ with $\frac{1}{\pi\omega}$, which can easily be evaluated. 

In order to extract the phase response of a system from spectral data through the Hilbert transform we therefore need a function where the phase and amplitude response are separate and individually linked to the real and imaginary parts of the function.  This can be found by taking the logarithm of the transmission coefficient $t(\omega)$

\begin{equation}
    t(\omega)=\sqrt{T(\omega)}e^{i\phi}
\end{equation}

Here $T=|t(\omega)|^2$ is the transmission and $\phi$ is the phase. The logarithm of $t(\omega)$ is now given by

\begin{equation}
    ln(t(\omega))=\frac{ln(T(\omega))}{2}+i\phi
\end{equation}

Hilbert transformations can only be applied if the relationship between transmission and phase is a causal one, i.e the transmission coefficient is a minimum phase function. While this is true for many metasurfaces it cannot always be assumed, e.g. it is not the case for polarisation altering metasurfaces, see ref \cite{5637099}.


For the metasurface used in this article, the transmission coefficient $t(\omega)$ is a minimum phase function. Hence Hilbert transforms are applicable and we get equation \ref{phi} for the linear phase of a metasurface.

\begin{equation}
    \phi(\omega_{0})=-\frac{1}{2\pi}\int_{-\infty}^{\infty} \frac{ln(T(\omega))}{\omega-\omega_{0}} d\omega.
    \label{phi}
\end{equation}

Nonlinear interactions can modify the properties of the metasurface, which we can express through an additional term introduced to the phase and transmission of the metasurface.

\begin{equation}
   \phi(\omega)+ \Delta \phi(\omega;\zeta) =-\frac{1}{2\pi}\int_{-\infty}^{\infty}\frac{ln(T(\omega)+\Delta T(\omega;\zeta))}{\omega-\omega_0}d\omega
   \label{NLeqn}
\end{equation}

Here $\Delta T$ is the change in transmission due to nonlinear optical effects and $\Delta \phi$ is the nonlinear phase change. In the case of slow nonlinear optical effects (slow compared to the time taken to perform a measurement) or nonlinear optical steady state measurements the terms $\Delta \phi$ and $\Delta T$ are constant over the duration of the measurement and such the Hilbert transform can be applied. (Note this might not apply to fast nonlinear optical effects or situations where a time-dependence of $\Delta \phi$ or $\Delta T$ has to be considered). 

Therefore eqn.\ref{phi} can be applied separately in the presence and absence of the nonlinear perturbation. The difference between the two reconstructed phase profiles corresponds to the nonlinear phase shift. We note that if the nonlinear effect on the transmission is to decrease it to close to zero in the high frequency (short wavelength) spectral ranges, then this results in high frequency noise in the reconstructed phase, significantly reducing the accuracy of the measurement. For typical metasurfaces we have significant transmission at all wavelength/frequencies and we are hence sufficiently far away from this restriction. 

\section{Simulation and Experimental Results}

\begin{figure}[h!]
    \centering
    \includegraphics[scale=0.45]{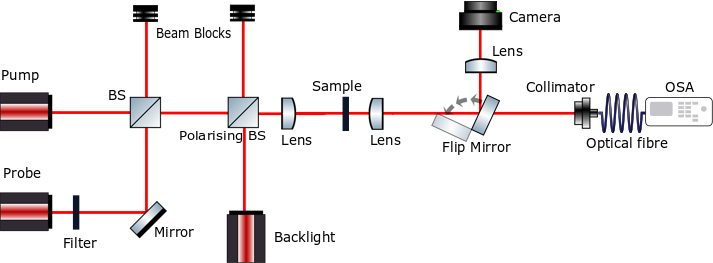}
    \caption{Sketch of the broadband pump-probe set-up. The pump laser is only switched on for the non-linear phase and transmission measurements.The first, nonpolarising beam splitter (BS), is used to combine the pump and probe beam. The polarising beam splitter is used to select the polarization parallel to the long axis of the antennas.}
    \label{Experimentsetup}
\end{figure}

\begin{figure}
    \centering
    \includegraphics[scale=0.35]{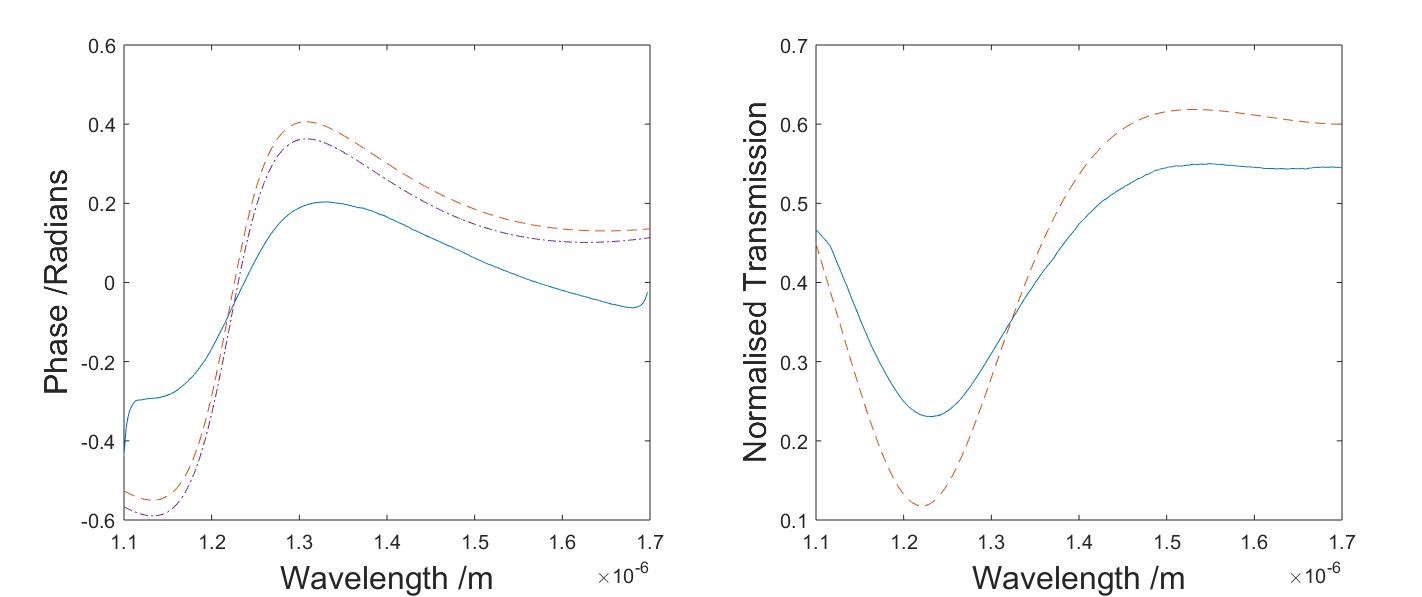}
    \caption{Left Panel: Linear phase response from Hilbert transform of the experimental transmission (solid blue line), FDTD S-parameter calculation (dashed red line) and Hilbert transform of the real FDTD transmission (dash-dotted purple line). Right Panel: Transmission data from experiment (solid blue line) and FDTD simulation (dashed red line).}
    \label{Comparison}
\end{figure}

Transmission measurements of the metasurface were taken using a SuperK Kompact broad-band source which was filtered using a high-pass filter to only include NIR wavelength ($\lambda\geq \SI{1.1}{\micro \meter}$), together with an Optical Spectrum Analyzer (OSA), set to record data at wavelengths between \SI{1.1}{\micro \meter} and \SI{1.7}{\micro \meter}, as shown in figure \ref{Experimentsetup}. The recorded transmission (see figure \ref{Comparison}) through the metasurface was normalised against the plain ITO on glass transmission of the sample before being smoothed using a third-order Savitzky-Golay filter \cite{doi:10.1063/1.4822961,MATLAB:2018}. The signal was then zero-padded and convolved with a Hanning window before the Hilbert transform was applied. Padding increases the signal span, yielding a more accurate result and the windowing eliminates the discontinuity introduced by the differences between the beginning and end of the signal, which would lead to unwanted artifacts in the transformed function.

FDTD simulations were performed to provide a comparison for the experimental data, using the commercial Lumerical FDTD solver \cite{Lumerical}. The antennas were modelled with average dimensions given by 397 nm by 151 nm by 40 nm, with a period of 600 nm, as collected from SEM images of the fabricated sample. The corners of the simulated antennas were rounded slightly to simulate the finite corner radius of the fabricated device. The polarisation of the incident light was parallel to the longest dimension of the antennas as in the experimental measurements. The ITO permittivity dispersion used in simulations was a Drude model fit with the epsilon-zero wavelength at 1470 nm. To extract the phase from the simulations we used two distinct approaches, using the Hilbert transform of the transmission spectra and, for comparison, calculating the S-parameters for the system. The S-parameters give the complex value of the zeroth-order transmission \cite{PhysRevE.71.036617,2010ITMTT..58.2646S}, hence the phase can be directly extracted by taking the angle of the complex transmission. The results of this analysis are shown in figure \ref{Comparison}. The right panel shows the experimental and numerical transmissions vs wavelength, whereas the left panel shows the extracted phase for the experimental case and for the two numerical approaches. The difference between the experimental and the numerical transmission curves are as follows: there is a very small shift in the minimum transmission wavelength and a slight broadening and weakening of the resonant feature. These differences can be attributed to slight differences between the antenna dimensions and the ITO epsilon-zero wavelength in simulations and experiment \cite{PhysRevA.93.063846}, but also to statistical variations in the antenna dimensions, variations in antenna corner rounding, film thickness variation, loss of antennas in the fabrication or chromatic features, i.e. a wavelength dependent misalignment between the metasurface and the broad band optical pump. All these factors lead to variations in the coupling between the metasurface and the incident beam and hence resonant features which differ in magnitude to those expected.

The weaker phase response observed experimentally is associated with the reduced strength of the experimentally observed optical resonance compared to the numerical resonance. The slight difference between the two FDTD results can be attributed to the fact that the signal used for the Hilbert transform has a finite spectral bandwidth. We note here that there is no free fitting parameter in the FDTD simulations, with all parameters (e.g. antenna dimensions) determined experimentally.

A typical design run of a specific meta-atom requires extensive and costly iterations and/or parameter sweeps. The computationally intensive nature of FDTD means that either processor time or memory requirements are typically the limiting factor. As seen in figure \ref{Comparison} both the simulated Hilbert transform phase and S-parameter phase are almost identical and could be used to predict the device behaviour. However, the S-parameter method required approximately 4.5GB of memory and consumed 68 processor hours. Followed by an additional 1h post-processing computation step, to extract the S-parameter and then phase from the recorded data. The Hilbert transform method on the other hand, only required the calculation of the transmission data, requiring 1.7GB of memory and 43 processor hours, with only seconds of post-processing computation. The numerical use of the Hilbert transform method therefore present a clear reduction in both the required memory and processor time, which over large parameter sweeps will significantly reduce simulation time, while allowing for potentially larger parallelization or use of lower performance (e.g. smaller memory, lower core number) computer systems. At the same time the experimental implementation results in a significantly simpler set-up, measurement and analysis compared to traditional interferometric set-ups.

\begin{figure}[h!]
    \centering
    \includegraphics[scale=0.3]{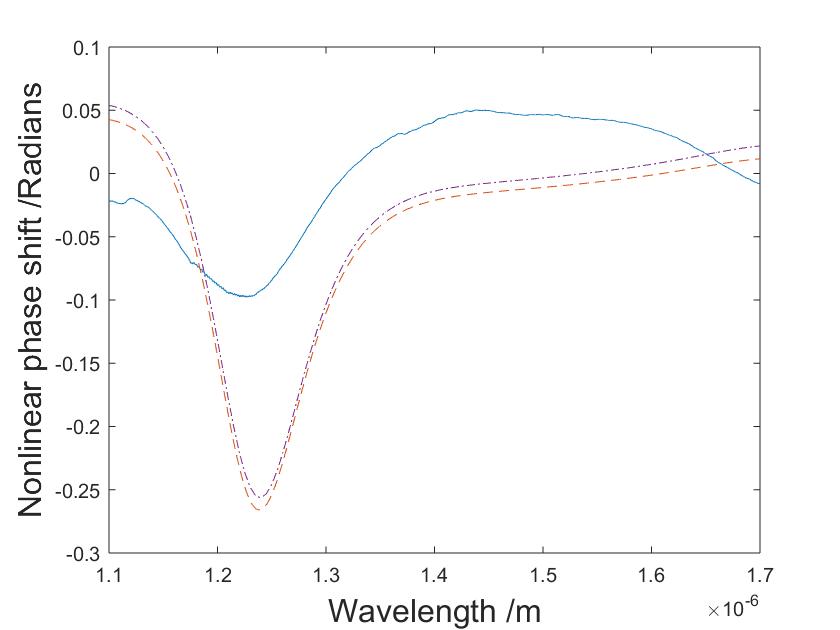}
    \caption{Nonlinear Phase shift from the Hilbert transform of the experimental data(solid blue line), Hilbert transform of FDTD transmission data(dash-dotted purple line),S-parameter calculation from FDTD(dashed red line).}
    \label{Nonlinear}
\end{figure}

Lastly, we address the use of Hilbert transform to extract a nonlinear phase shift. Within this experiment we used a thermal nonlinearity, which will result in a shift of the ITO dispersion curve, which in turn will alter the wavelength response of the system. To simulate the nonlinearity, the dispersion for the ITO was manually increased and decreased by different percentages and sweeps using the altered permittivities were run until the simulations gave the closest match to the resonance position of the experimental data. This simulates the change in plasma frequency brought about by pumping the sample. In the experiment, the metasurface was pumped using a Yenista optics Tunics T100S-HP laser set to a power of 0.5mW and a wavelength of 1550 nm.  A 25\% increase in the relative permittivities gives the shift in resonance features observed in the experimental results. The optical phases (experiment and simulations) were then extracted using the same methods as earlier. By subtracting away the linear phase from the phase found for the sample under pump irradiance the nonlinear phase shift, $\Delta\phi$ can be found, see figure \ref{Nonlinear}.

We note that once again the S-parameter and Hilbert transform methods yield almost identical phase responses, validating the use of the Hilbert method for this approach and confirming that the systems transmissivity still fulfils the conditions of causality and being a linear equation. The difference between experimental and simulated data once again reflects the weaker response observed in experiment. However, the key features, e.g. resonance position and overall shape are reproduced. 
The Hilbert transform method for nonlinear phase extraction has many advantages compared to alternative approaches. The same issue of interferometric stability, and sensitivity to external variations that affect linear FTSI are also present during nonlinear FTSI, potentially amplified by the increased measurement time needed to for example, sweep pump power. Using Hilbert transformations, nonlinear phase shifts can be evaluated over a broad spectral range rather than being limited to a single frequency as would be the case in z-scan measurements. Z-scan measurements would be inappropriate in our case anyway, as the set-up depends on focusing and defocusing of light by a sample due to a variance in incident intensity, while the thermal nonlinearity here varies with power per unit volume as opposed to intensity. We note that the Hilbert transform method is applicable as long as the nonlinear process can be seen as a steady state response, compatible with \ref{NLeqn}

\section{Summary}

In this paper we have demonstrated that Hilbert transforms can be applied to transmission data to extract broadband linear and nonlinear phase shifts of optical metasurfaces. Compared to the traditional FTSI method we have reduced the need for interferometric set-ups and stability, making measurements easier and more robust to external conditions. We also note that the Hilbert transform can work with lower sampling resolution compared to the FTSI, allowing for the use of lower resolution spectrometers, unless the transmission spectra of the metasurfaces have spectrally narrow features. Furthermore, we have shown that the Hilbert method has advantages in phase gradient metasurface design, significantly reducing the memory and processor requirements needed to extract the design phase compared to the typical S-parameter calculations. We have also demonstrated the applicability of the Hilbert transform to the measurement of thermo-optic nonlinear phase shifts. The method presented here shall be a particularly useful tool in the future design and characterisation of dynamically tunable metasurfaces.   

\vspace{1cm}

\section{Acknowledgements:}
This work is funded through the DASA "Metasurfaces for defence and security" grant ACC6004047. Laura Wynne is funded through an EPSRC DTA scholarship. Hamish Ballantyne acknowledges funding through the Royal Society Research Grant RGS$\setminus$R2$\setminus$180054. We acknowedge useful discussions with Dr Jeremy Upham (University of Ottawa).
The data underlying this work will be made available (open access) online after publication.

\bibliographystyle{elsarticle-num}
\bibliography{main}

\end{document}